\documentclass[10pt]{article}
\usepackage{graphicx}

\usepackage{amsfonts}
\usepackage{longtable}
\usepackage[toc,page]{appendix}
\usepackage{amsmath}
\usepackage{amssymb}
\usepackage{eufrak}
\usepackage{epstopdf}
\usepackage{color}

\usepackage[colorlinks, citecolor=blue, linkcolor=blue, backref=true]{hyperref}
\usepackage{longtable}
\usepackage{url}

\usepackage[firstinits=true, sorting=none,parentracker=true, doi=true, isbn=false, natbib=true]{biblatex}
\usepackage{upgreek}

\usepackage{amsmath,environ}
\NewEnviron{eqn}{\begin{equation}\begin{split}
 \BODY
\end{split}\end{equation}
}
\newcommand{\beqans}{\begin{subequations}\begin{eqnarray}}
\newcommand{\eeqans}[1]{\end{eqnarray}\label{#1}\end{subequations}}
\newcommand{\chinew}{\chi}

\usepackage{mathrsfs}

\DeclareMathAlphabet{\mathpzc}{OT1}{pzc}{m}{it}
\newcommand{\KK}{\mathtt{K}}
\newcommand{\elltg}{\mathfrak{t}}
\newcommand{\ellhx}{\mathfrak{h}}
\newcommand{\su}{u}
\newcommand{\dxp}{\mathtt{x}}
\newcommand{\dyp}{\mathtt{y}}
\newcommand{\sv}{\mathtt{v}}
\newcommand{\phasefac}{\uppsi}

\begin{document}
\title{Wiener--Hopf factorisation on unit circle: some examples from discrete scattering}
\author{Basant Lal Sharma\thanks{Department of Mechanical Engineering, Indian Institute of Technology Kanpur, Kanpur, U. P. 208016, India ({bls@iitk.ac.in})}}
\maketitle

\begin{abstract}
{I discuss some problems featuring scattering due to discrete edges on certain structures. These problems stem from linear difference equations and the underlying basic issue can be mapped to Wiener--Hopf factorization on an annulus in the complex plane. In most of these problems, the relevant factorization involves a scalar function, while in some cases a $n\times n$ matrix kernel, with $n\ge2$, appears. For the latter, I give examples of two non-trivial cases where it can be further reduced to a scalar problem but in general this is not the case. Some of the problems that I have presented in this paper can be also interpreted as discrete analogues of well-known scattering problems, notably a few of which are still open, in Wiener--Hopf factorization on an infinite strip in complex plane.}
\end{abstract}

\section*{Introduction}
\label{intro}
The factorization technique conceived by Wiener and Hopf \cite{Wiener}, in a strip or an annulus in the complex plane \cite{Krein}, makes possible the construction of an additive/multiplicative decomposition of analytic functions that has wide applications in science and engineering. In particular, it has been a useful tool to resolve successfully several outstanding issues arising in the theory of scattering of waves in physical systems \cite{Bouwkamp,Vino}. From a technical viewpoint, its extension from the class of scalar functions to that of matrix-valued functions \cite{HeinsSys,KreinGoh}, has been found to be challenging and non-trivial. The subject contains several open problems including the need of constructive methods \cite{Rogosin2016}. In this paper, I present certain examples of lattice structures which involve scalar, in general matrix, Wiener--Hopf ({{WH}}) factorisation. Special cases of two examples are presented with matrix kernels which are equivalent to a scalar problem via an original construction.

\begin{figure}[h!]
\centering
\includegraphics[width=\textwidth]{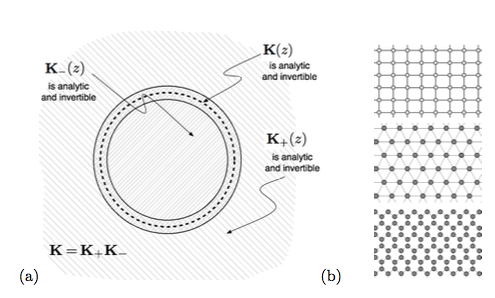}
\caption{(a) Multiplicative {{WH}} factorization of a $n\times n$ matrix-valued function
($n=1$ corresponds to scalar case) which is analytic on an annulus in complex plane (dashed curve is the unit circle). 
(b) {Three kinds of lattice structures discussed in this paper, namely, square, triangular, and honeycomb.}}
\label{annuluscomplexplane}
\end{figure}

The program of this paper is realized by identifying and posing certain scattering problems \cite{Felsen}
on lattice structures \cite{Brillouin}.
The relevant question of factorization is illustrated schematically in Fig. \ref{annuluscomplexplane}(a).
It is required that $\mathbf{K}({z})=\mathbf{K}_+({z})\mathbf{K}_-({z})$ for ${z}$ in an annulus ${\mathcal{A}}$ in the complex plane, where each of matrix functions $\mathbf{K}$, $\mathbf{K}_+$, $\mathbf{K}_-$ is analytic and invertible on the annulus while $\mathbf{K}_+$ (resp. $\mathbf{K}_-$) is also analytic and invertible outside (resp. inside) it.
A schematic of the underlying three lattice structures is shown in Fig. \ref{annuluscomplexplane}(b).
It is worth pointing out that, as a departure from the more popular continuum formulation, the discrete models have been playing an important role in understanding the mechanics of singularities and critical processes; the latter occurs partly due to a regularizing effect as well as a need to capture physical effect associated with the atomic resolution of structures \cite{thomson1971lattice,disloc_slepyan1,Marderatom,disloc_Thomson,novoselov2005two_2}. 
Indeed, the same applies to the continuum theory of scattering also, though sometimes the terminology refers to the relevant difference operators as discrete Schr\"{o}dinger operators \cite{Korotyaev2014,isozaki2012inverse}.
Remarkably, the {\em discrete} problem of scattering by defects in arbitrary lattices is endowed with several distinguished contributions \cite{Lifshitz1948,Maradudinbook}; many of these were contemplated decades ago. On the other hand, I have recently explored the concomitant issues dealing with the propagation of waves interacting with edges of cracks and rigid inclusions in a series of papers \cite{Bls0,Bls1,Bls2,Bls3,Bls4,Bls5,Bls6,Bls9s,Bls8pair1,Bls8staggerpair_asymp,Bls10mixed}.
In these papers, a fundamental role is played by the classical Sommerfeld half-plane \cite{Sommerfeld1}; Sommerfeld problem, which involves the two dimensional Helmholtz equation with either Dirichlet boundary condition or Neumann boundary condition on a half-line, arises in the phenomena originating due to electro-magnetic, acoustic, or elastic waves \cite{Vino, Achenbach}. 

In this paper, I present a partial review of certain discrete scattering problems, while
including a discussion of some of their novel aspects as well as stating a new set of unsolved problems. 
In \S\ref{scalarprobs}, I present the examples of discrete scattering problems following discrete analogues of both Sommerfeld problems on three structures shown in Fig. \ref{annuluscomplexplane}(b); only a brief formulation of the discrete {{WH}} equation \cite{Wiener, Karp, Feld,Noble} is provided using the discrete Fourier transform. 
In \S\ref{matrixprobs1}, I present certain problems that involve a $2\times2$ matrix kernel; these problems can be reduced to scalar problem(s) by using either certain symmetry in the physical structure or by seeking certain symmetry through a choice of alternative coordinates. The latter is omitted in this paper for the triangular and honeycomb lattices, as the details of an approach based on an alternative choice of co-ordinates can be found in \cite{Bls4,Bls5,Bls6}. 
In \S\ref{matrixprobs3}, I include the derivation of a hierarchy of matrix kernels, of arbitrary order;
more popular via certain continuum scattering problems \cite{AbrahamsExpo,Abrahams0,Abrahams2,Abrahams4}. These matrix kernels arise in the theory of discrete scattering due to a finite array of, typically staggered, cracks and rigid constraints. This includes a special case of $2\times2$ kernels as well, which I have recently presented elsewhere \cite{Bls8staggerpair_exact,Bls32} wherein it is shown that the problem can be reduced to a scalar problem and an auxiliary linear algebraic equation.
In the same section, I provide a glimpse of discrete scattering problems with oppositely oriented pair of cracks and rigid constraints
whose matrix kernels have similar structure; from another perspective as discrete analogues \cite{Luneburg1,Rawlins469,HurdLune}. Using the same scaling, or similar, as utilized in \cite{Bls31,Bls2}, a rigorous limit to continuum exists, however, this is omitted here.
In \S\ref{discussn}, I conclude this paper with some remarks. 

\section{Examples of discrete scattering with scalar {{WH}} factorization}
\label{scalarprobs}
In this section I review some recently solved problems of discrete scattering on three kinds of lattice structures (Fig. \ref{annuluscomplexplane}(b)) whose solution depends on a discrete {{WH}} equation with a scalar kernel.
The first problem concerns scattering due to a semi-infinite static crack in square lattice \cite{Bls0} while the second problem relates to atomically sharp rigid inclusions \cite{Bls1}. The third and fourth problems are based on triangular and honeycomb lattices \cite{Bls4,Bls5}.
The discrete Neumann conditions follow a model originally postulated for a moving crack in square lattice \cite{disloc_slepyan1,Slepyanbook}. Throughout the paper, $\su^{{t}}_{x,y}$ denotes total (out-of-plane) displacement field at lattice site $(x,y)$. The additive {{WH}} factors are denoted by {\em super}script $\pm$ while multiplicative ones by {\em sub}script $\pm$, i.e., for a suitable function $f$, $f=f^++f^-$, $f=f_+f_-$. ${\mathbb{C}}$ is the set of complex numbers,
${\mathbb{Z}}$ is the set of integers, ${\mathbb{Z}^+}$ (resp. ${\mathbb{Z}^-}$) is the set of non-negative (resp. negative) integers.

\subsection{}
\begin{figure}[h!]
\centering
\includegraphics[width=\textwidth]{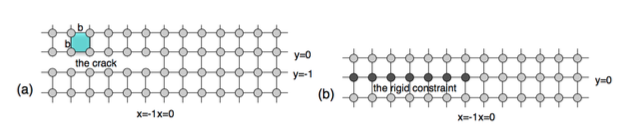}
\caption{Square lattice with a semi-infinite row prescribed with discrete (a) Neumann (crack) \cite{Bls0,Bls2}, (b) Dirichlet (rigid constraint) \cite{Bls1,Bls3} condition. The cell shaded with green color is the unit cell.}
\label{sqdiffracK}
\end{figure}

\begin{figure}[h!]
\centering
\includegraphics[width=\textwidth]{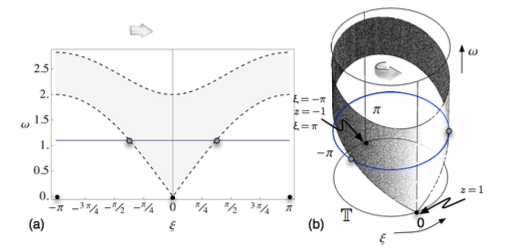}
\caption{(a) An illustration of pass band (based on dispersion relation). (b) Result of mapping of (a) using real wave numbers ${\upxi}\in[-\pi, \pi]$ to complex numbers via ${z}=e^{-i{\upxi}}$ (on unit circle) in complex plane.}
\label{sqlattdispersion}
\end{figure}

Consider the structure shown schematically in Fig. \ref{sqdiffracK}(a). 
The equation of motion of each particle in the square lattice (with spacing $b$), away from the crack faces, is assumed to be described by $M\frac{d^2}{dt^2}{\su}_{\dxp, \dyp}(t)={K}({\su}_{\dxp+1, \dyp}(t)+{\su}_{\dxp-1, \dyp}(t)+{\su}_{\dxp, \dyp+1}(t)+{\su}_{\dxp, \dyp-1}(t)-4{\su}_{\dxp, \dyp}(t))/{b^2}.$
In a constant frequency scenario, for instance, in a steady state description of the lattice dynamics in the presence of an excitation due to incident wave field of certain frequency, omitting the time dependence and the factor $e^{-i\omega t}$, the equation of motion reduces to an algebraic equation, namely, a linear partial difference equation in two independent variables $\dxp$ and $\dyp$ \cite{levy}. This is applicable throughout the paper as well. It is assumed that the lattice is excited by an (out-of-plane) incident wave with a spatial part given by
\begin{eqn}
{\su}_{\dxp, \dyp}^{\mathrm{in}}{:=}{{\mathrm{A}}}e^{-i{\upkappa}_x \dxp-i{\upkappa}_y \dyp}, 
\label{uinc}
\end{eqn}
while the total displacement field ${\su}^{{t}}$, away from crack faces, satisfies discrete Helmholtz equation 
\begin{eqn}
{\su}^{{t}}_{\dxp+1, \dyp}+{\su}^{{t}}_{\dxp-1, \dyp}+{\su}^{{t}}_{\dxp, \dyp+1}+{\su}^{{t}}_{\dxp, \dyp-1}-4{\su}^{{t}}_{\dxp, \dyp}+\upomega^2{\su}^{{t}}_{\dxp, \dyp}=0
 \text{ with } \upomega=\omega{b}\sqrt{M/K}.
 \label{dHelmholtz}
 \end{eqn}
Throughout the paper, $\su_{\dxp,\dyp}$ denotes the scattered wave field at lattice site $(\dxp,\dyp)$,
\begin{eqn}
{\su}_{\dxp, \dyp}={\su}_{\dxp, \dyp}^{{t}}-{\su}_{\dxp, \dyp}^{\mathrm{in}}
\end{eqn}
Notice that given $\upomega$, the incident wave number ${\upkappa}_x$ and ${\upkappa}_y$ are related by the square lattice 
dispersion relation $\upomega^2=2-2\cos{\upkappa}_x-2\cos{\upkappa}_y$. The real values of ${\upomega}, {\upkappa}_x$ ($={\upxi}$) for real ${\upkappa}_y$ correspond to the pass band as shown in grey color schematically in left part of Fig. \ref{sqlattdispersion} (the blue line corresponds to a fixed ${\upomega}$).
To attend the issue of causality and well-posedness of certain features of the mathematical formulation \cite{Noble}, it is further assumed that ${\upomega}={\upomega}_1+i{\upomega}_2\in{\mathbb{C}}$ with ${\upomega}_2>0$; thus, issue of radiation condition \cite{Shaban} is made convenient.
In the presence of positive imaginary part of ${\upomega}$, the circular contour can be mapped to a contour in the annulus of Fig. \ref{annuluscomplexplane}(a) while the intersection between the blue curve and gray region maps to the branch cuts with pieces lying inside or outside the annulus \cite{Bls0}.
Since the incident wave ${\su}_{\dxp, \dyp}^{\mathrm{in}}$ \eqref{uinc} satisfies discrete Helmholtz equation \eqref{dHelmholtz}, it is seen that the scattered field ${\su}_{\dxp, \dyp}$ also satisfies the same equation and is required to decay away from the crack faces. Taking discrete Fourier transform of \eqref{dHelmholtz} for ${\su}_{\dxp, \dyp}$ along the $\dxp$ coordinate according to the definition
\begin{eqn}
{\su}_{\dyp}^{\mathrm{F}}({{z}})={\su}_{\dyp}^{+}({{z}})+{\su}_{\dyp}^{-}({{z}}), {\su}_{\dyp}^{+}({{z}})=\sum\nolimits_{\dxp\in{\mathbb{Z}^+}} {\su}_{\dxp, \dyp}{{z}}^{-\dxp}, {\su}_{\dyp}^{-}({{z}})=\sum\nolimits_{\dxp\in{\mathbb{Z}^-}} {\su}_{\dxp, \dyp}{{z}}^{-\dxp},
\label{unpm}
\end{eqn}
the scattered wave field is found to be described by \cite{Bls0}
\begin{eqn}
{\su}_{\dyp}^{\mathrm{F}}({{z}})=
{\su}_0^{\mathrm{F}}({{z}}){{\lambda}}^{\dyp}({{z}}), \dyp \ge 0, \forall {{z}}\in{{\mathcal{A}}}.
\label{ubulk}
\end{eqn}
where (owing to \eqref{dHelmholtz} and \eqref{ubulk}, ${\lambda}+{\lambda}^{-1}+{z}+{z}^{-1}-4+{\upomega}^2=0$) 
\begin{eqn}
{{\lambda}}({{z}}){:=}\frac{\mathpzc{r}({{z}})-\mathpzc{h}({{z}})}{\mathpzc{r}({{z}})+\mathpzc{h}({{z}})}, {\mathpzc{h}}({{z}}){:=}\sqrt{2-{z}-{z}^{-1}-{\upomega}^2}, {\mathpzc{r}}({{z}}){:=}\sqrt{{\mathpzc{h}}^2({{z}})+4}, 
 {{z}}\in{\mathbb{C}}\setminus{\mathscr{B}},
 \label{lam12}
\end{eqn}
and ${\mathscr{B}}$ denotes the branch cut for ${{\lambda}}$, resulting from the chosen branch for ${\mathpzc{h}}$ and ${\mathpzc{r}}$ \cite{Slepyanbook},
such that 
$|{{\lambda}}({{z}})|\le1, {{z}}\in{\mathbb{C}}\setminus{\mathscr{B}}.$
Throughout this paper, the branch cut of complex function $\sqrt{.}$ is chosen to coincide with the negative real axis.
The annulus ${{\mathcal{A}}}$ mentioned in \eqref{ubulk} is found \cite{Bls0} to be contained in 
$\{{{z}}\in{\mathbb{C}}: e^{-{\upkappa}_2}< |{{z}}|< e^{{\upkappa}_2\cos{\Theta}}\},$
where ${\upkappa}_2$ is the (positive) imaginary part of $\sqrt{{\upkappa}_x^2+{\upkappa}_y^2}$ while ${\Theta}$ is the incidence angle relative to the positive $\dxp$ axis.
In the context of the pass band of the lattice, a visualization of the annulus ${{\mathcal{A}}}$, which almost coincides with the unit circle as ${\upomega}_2\to0$, is provided in the right part of Fig. \ref{sqlattdispersion}. At this point, it is also pertinent to recollect Fig. 2 from \cite{Bls0} as well as \cite{Bls2}.
In the remainder of this paper, the discrete scattering problems have been formulated {\em without} the provision of a detailed description of the annulus ${{\mathcal{A}}}$ in terms of the incident wave parameters and specific lattice structure; the details can be found in \cite{Bls0,Bls3,Bls6}.
Since the incident wave \eqref{uinc} does not fulfil the condition of absence of bonds across the crack faces, it is clear that the broken bonds across the crack faces act as sources, for the scattered wave field, which are derived from the incident wave. These sources have an equal and opposite sign across the semi-infinite rows of particles with $\dxp<0$ for $\dyp=0$ vs $\dyp=-1$; as a consequence, the scattered wave field is odd-symmetric across mid-point of $\dyp=0$ and $\dyp=-1.$ 
In particular, by \eqref{ubulk}, it is sufficient to solve for $\su_0^{\mathrm{F}}$, precisely the objective of the following equation.
Taking the Fourier transform \eqref{unpm} along the lattice rows $\dyp=0$ 
(counterpart of the discrete Helmholtz equation \eqref{dHelmholtz}, ${\su}^{{t}}_{\dxp+1, 0}+{\su}^{{t}}_{\dxp-1, 0}+{\su}^{{t}}_{\dxp, 1}+({\upomega}^2-3){\su}^{{t}}_{\dxp, 0}=0$ for $\dxp<0$ and for $\dxp\ge0$ \eqref{dHelmholtz} holds with $\su_{\dxp, -1}=-\su_{\dxp, 0}$)
it is found that 
the {{WH}} equation to obtain $\su_0^{\mathrm{F}}$ is
\begin{eqn}
f^{+}({z})+\KK({z})f^{-}({z})=c({z}), {z}\in{\mathcal{A}},\label{scalarWHeq}
\end{eqn}
\begin{eqn}
\text{where }
{{\KK}}_{{}}({{z}}){:=}\frac{{\mathpzc{h}}({{z}})}{{\mathpzc{r}}({{z}})}=\frac{1-{\lambda}({{z}})}{1+{\lambda}({{z}})},
f^\pm({{z}})={\su}_{0}^{\pm}({{z}}), 
c({z})=\tfrac{1}{2} (1-{{\KK}}_{{}}({{z}}))({\su}^{\mathrm{in}}_{0}{}^{-}({{z}})-{\su}^{\mathrm{in}}_{-1}{}^{-}({{z}})).
 \label{WHeqBls0}
\end{eqn}

Now I discuss the second problem as a discrete analogue of the Dirichlet boundary condition on the Sommerfeld plane \cite{Sommerfeld1}, namely,
the problem of scattering due to a semi-infinite rigid constraint in square lattice \cite{Bls1}. In this case $\su^{{t}}_{\dxp,0}=0, \dxp<0$; see Fig. \ref{sqdiffracK}(b) for a schematic illustration of the lattice structure.
The manipulations leading to the expression of the scattered wavefield are similar to that for the crack except that the field is even-symmetric about $\dyp=0$, in particular, ${\su}_{\dxp, 1}={\su}_{\dxp, -1}$. 
The {{WH}} equation \eqref{scalarWHeq} re-occurs with \eqref{WHeqBls0} replaced by
\begin{eqn}
{{\KK}_{{}}}{:=}\frac{{\mathpzc{h}}^2+2}{{\mathpzc{r}}{\mathpzc{h}}}=\frac{1+{\lambda}^2}{1-{\lambda}^2}, 
f^{\pm}={\su}_{1}^{\pm}, 
c({z})=\tfrac{1}{2}(1-{{\KK}_{{}}}({z}))((\mathpzc{h}^2({z})+2){\su}^{\mathrm{in}}_{0}{}^{-}({z})+{\su}^{\mathrm{in}}_{-1, 0}+{{z}} {\su}_{0, 0}).
\label{WHeqBls1}
\end{eqn}
From the viewpoint of the standard application of the {{WH}} technique \cite{Noble}, it is noted that in \eqref{scalarWHeq} $f^+$ (resp. $f^-$) is analytic on the annulus ${\mathcal{A}}$ and outside (resp. inside) it.
The {{WH}} kernel $\KK$ for both problems \cite{Bls0,Bls1} admits a multiplicative factorization $\KK=\KK_+\KK_-$ in closed form and the solution can be constructed neatly. Moreover, the closed form expressions for the lattice sites near the crack tip and the constraint tip can be also found \cite{Bls2,Bls3}.

Besides the two simple cases summarized above, there is a one-parameter family of discrete scattering problems for each of the two problems \eqref{WHeqBls0} and \eqref{WHeqBls1} which is obtained by relaxing the condition of broken bonds to merely bonds with different stiffness $c$ and replacing the rigid constraint by particles of mass $m$; $c$ and $m$ are clearly the relevant parameters. It is evident that the resulting problems can be analyzed in the same manner as presented in \cite{Bls0,Bls1,Bls2,Bls3} however the {{WH}} kernels do not admit closed form expressions for multiplicative factorization for arbitrary choice of $c$ and $m$. Some work in this direction is currently under investigation and is anticipated to appear elsewhere \cite{MishurisP}.
\begin{figure}[h!]
\centering
\includegraphics[width=\textwidth]{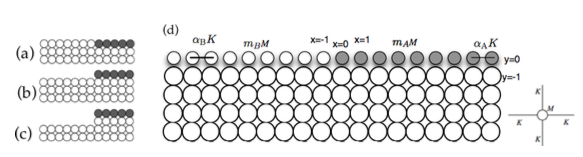}
\caption{(a)--(c) {Semi-infinite step/mixed boundary} \cite{Bls10mixed}, (d) structured boundary in half-plane \cite{Victor_Bls_surf2}.}
\label{threesteps_interface}
\end{figure}

Also, there are several other closely related problems to the above two discrete scattering problems on infinite square lattice. For instance, those presented by \cite{Bls10mixed} concern
the analysis of scattering due to a step/mixed boundary in semi-infinite lattices;
see Fig. \ref{threesteps_interface}(a)--(c) for a schematic. These problems can be reduced to the {{WH}} equation of type \eqref{scalarWHeq} though the multiplicative decomposition of {{WH}} kernel does not admit closed form expression.
Similar scalar {{WH}} equation \eqref{scalarWHeq} also arises in the study of surface wave propagation \cite{Victor_Bls_surf1}
across a structured boundary interface in semi-infinite lattice \cite{Victor_Bls_surf2};
see Fig. \ref{threesteps_interface}(d) for an illustration of the geometry. 
On the other hand, upon replacing the square lattice by other lattices, such as triangular lattice and honeycomb lattice, yields another pair of discrete analogues of Sommerfeld scattering problems \cite{Sommerfeld1}. In particular, for the discrete Dirichlet condition on a half-row of triangular lattice \cite{Bls4} and discrete Neumann condition \cite{Bls5} on a half-row of zigzag \cite{Nakada1996_1,Acik2011} honeycomb lattice, I briefly discuss below how the {{WH}} equation of type \eqref{scalarWHeq} occurs.

\subsection{}
\begin{figure}[h!]
\centering
\includegraphics[width=\textwidth]{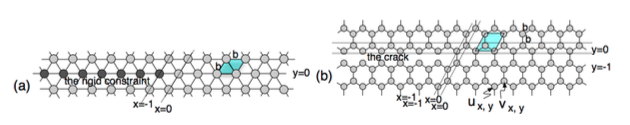}
\caption{Triangular lattice with a semi-infinite row prescribed with discrete Dirichlet (rigid constraint) condition \cite{Bls4,Bls6}, (b) Honeycomb lattice with a semi-infinite row prescribed with discrete Neumann (crack) \cite{Bls5,Bls6} condition.
The cell shaded with green color is the unit cell.}
\label{tghxdiffracCK}
\end{figure}

I consider now the {{WH}} formulation for discrete scattering of incident wave \eqref{uinc} associated with a semi-infinite rigid constraint in triangular lattice \cite{Bls4}; see Fig. \ref{tghxdiffracCK}(a) for an illustration of the structure and (slant) lattice coordinates. 
Analogous to \eqref{dHelmholtz} for square lattice, the discrete Helmholtz equation for the total wave field on the triangular lattice is 
a linear partial difference equation for ${u}^{{t}}$, i.e.,
\begin{eqn}
{u}^{{t}}_{\dxp+1, \dyp}+{u}^{{t}}_{\dxp-1, \dyp}+{u}^{{t}}_{\dxp, \dyp+1}+{u}^{{t}}_{\dxp, \dyp-1}+{u}^{{t}}_{\dxp-1, \dyp+1}+{u}^{{t}}_{\dxp+1, \dyp-1}
-6{u}^{{t}}_{\dxp, \dyp}+\tfrac{3}{2}{\upomega}^2{u}^{{t}}_{\dxp, \dyp}=0.
\label{discHelmtriang}
\end{eqn}
By inspection of Fig. \ref{tghxdiffracCK}(a), there is a visible even-symmetry about the rigid constraint. Naturally, it turns out that the {{WH}} formulation resembles that for the square lattice. It is easy to see that the scattered wave field satisfies $u_{\dxp, \dyp}=u_{\dxp+\dyp, -\dyp}, \dyp\ge0$ according to the choice of coordinates used in Fig. \ref{tghxdiffracCK}(a).
Taking the Fourier transform \eqref{unpm} and assuming that ${u}_{\dyp}^{\mathrm{F}}({z})\to 0$ when ${\dyp}\to\infty$, the solution of the discrete Helmholtz equation \eqref{discHelmtriang} for $\su$ is expressed as
\begin{eqn}
{u}_{\dyp}^{\mathrm{F}}({z})=u_0^{\mathrm{F}}({z})\elltg^{{\dyp}}({z}),
{\dyp}>0,
\label{disctriang10}
\end{eqn}
\begin{eqn}
\text{where }\elltg^2-F({z})\elltg+{z}=0\text{ with }F({z})=(6-{z}-{z}^{-1}-\tfrac{3}{2}{\upomega}^2)/(1+{{z}}^{-1}),
\label{disctrianglambda}
\end{eqn}
such that $|\elltg|<1$ in the annulus ${\mathcal{A}}$.
When ${\dyp}=0$ and
$\dxp<0,$
the discrete Helmholtz equation \eqref{discHelmtriang} is replaced by
${u}^{{t}}_{\dxp, 0}=0$ whereas it remains intact for $\dxp\ge0$ except that $u_{\dxp,-1}=u_{\dxp-1,1}, u_{\dxp+1, -1}=u_{\dxp,1}$.
As a result of \eqref{disctriang10},
${u}_{1}^{+}+{u}_{1}^{-}=({u}_{0}^{+}+{u}_{0}^{-})\elltg$ leads to
the {{WH}} equation \eqref{scalarWHeq}, which decides $u_0^{\mathrm{F}}({z})$ in \eqref{disctriang10} for ${z}\in{\mathcal{A}}$, with \begin{eqn}
{{\KK}_{{}}}{:=}
\frac{F}{F-2\elltg}, 
f^\pm={\su}_{1}^{\pm}, 
c({z})=\tfrac{1}{2}(1-{{\KK}_{{}}}({{z}}))(F({z}){\su}^{\mathrm{in}}_{0}{}^{-}({z})+\frac{{\su}^{\mathrm{in}}_{-1, 0}-2u_{-1,1}+{{z}} {\su}_{0, 0}}{1+{z}^{-1}}).
\label{WHeqBls4c}
\end{eqn}

I present now another instance of a discrete scattering problem that possesses a symmetry that eventually leads to a 
scalar {{WH}} problem. This is the case of scattering of out-of-plane waves due to a semi-infinite crack in zigzag honeycomb lattice \cite{Bls5}; the potential applications involving the geometry of honeycomb structure are well-known \cite{Novoselov,geim2007rise}. See Fig. \ref{tghxdiffracCK}(b) for an illustration of the structure and choice of coordinates (following the slant coordinates for triangular lattice).
In this case, the equations of motion that are postulated to be solved for the total wave field $u^{{t}}, v^{{t}}$ (a pair due to the fact that it is not a simple lattice) are
\beqans
{v}^{{t}}_{\dxp, {\dyp}}+{v}^{{t}}_{\dxp-1, {\dyp}}+{v}^{{t}}_{\dxp, {\dyp}-1}-3{u}^{{t}}_{\dxp, {\dyp}}+\tfrac{3}{4}{\upomega}^2{u}^{{t}}_{\dxp, {\dyp}}&=&0, \label{dHelmholtzhx1}\\
{u}^{{t}}_{\dxp, {\dyp}}+{u}^{{t}}_{\dxp+1, {\dyp}}+{u}^{{t}}_{\dxp, {\dyp}+1}-3{v}^{{t}}_{\dxp, {\dyp}}+\tfrac{3}{4}{\upomega}^2{v}^{{t}}_{\dxp, {\dyp}}&=&0.
\label{dHelmholtzhx2}
\eeqans{dHelmholtzhx}
The incident wave field comprises of a pair, ${u}_{\dxp, {\dyp}}^{\mathrm{in}}$ and ${v}_{\dxp, {\dyp}}^{\mathrm{in}},$ which is assumed to satisfy \eqref{dHelmholtzhx} for given ${\upomega}$; the expression of ${u}^{\mathrm{in}},{v}^{\mathrm{in}}$ is analogous to \eqref{uinc}. It is seen that the scattered field described by ${u}_{\dxp, {\dyp}}$ and ${v}_{\dxp, {\dyp}}$ also satisfies the same equation \eqref{dHelmholtzhx}; moreover, ${u}$ and ${v}$ are required to decay away from the crack faces.
Similar to case of crack in square lattice, it is clear that the broken bonds across particles located at crack faces act as sources, derived from the incident wave, with an equal and opposite sign so that the scattered wave field is odd-symmetric;
in particular, 
$u_{\dxp, 0}=-v_{\dxp, -1}$ for $\dxp<0$.
Taking the Fourier transform \eqref{unpm} of \eqref{dHelmholtzhx} for $u, v$ and assuming that ${u}_{\dyp}^{\mathrm{F}}({z})\to 0$, ${v}_{{\dyp}}^{\mathrm{F}}({z})\to 0$ when ${\dyp}\to\infty$, 
it is found that
\begin{eqn}
{u}_{\dyp}^{\mathrm{F}}({z})=u_0^{\mathrm{F}}({z})\ellhx^{{\dyp}}({z}),
{v}_{{\dyp}}^{\mathrm{F}}({z})=
u_0^{\mathrm{F}}({z})\frac{1+z+\ellhx}{3(1-\tfrac{1}{4}{\upomega}^2)}\ellhx^{{\dyp}}({z}), {\dyp}>0,
\label{dischxsol}
\end{eqn}
where $\ellhx$ is given by the solution of an equation almost same as \eqref{disctrianglambda}, namely, $\ellhx^2-F({z})\ellhx+{z}=0$, with $F({z})=(6-{z}-{z}^{-1}-\tfrac{3}{2}{\upomega_{T}}^2)/(1+{{z}}^{-1})$ and 
${{\upomega_{T}}}^2=\tfrac{3}{2}{\upomega}^2(2-\tfrac{1}{4}{\upomega}^2),$
such that $|\ellhx|<1$ in the annulus ${\mathcal{A}}$. 
Taking the Fourier transform \eqref{unpm} along the lattice rows ${\dyp}=0$ 
(with the counterpart of \eqref{dHelmholtz}, $-\tfrac{3}{4}{\upomega}^2{u}^{{t}}_{\dxp, {\dyp}}={v}^{{t}}_{\dxp, {\dyp}}+{v}^{{t}}_{\dxp-1, {\dyp}}-2{u}^{{t}}_{\dxp, {\dyp}}$ for $\dxp<0$ and for $\dxp\ge0$ \eqref{dHelmholtz} holds with $v_{\dxp, -1}=-u_{\dxp, 0}$)
it is found that the discrete {{WH}} equation \eqref{scalarWHeq} needs to be solved for ${\su}_0^{\mathrm{F}}$, the only unknown in \eqref{dischxsol}, 
with
\begin{eqn}
{{\KK}}_{{}}=\frac{{{\mathpzc{N}}}-1}{{{\mathpzc{N}}}+1},f^\pm={\su}_{0}^{\pm}, 
c=\tfrac{1}{2}(1-{{\KK}_{{}}})({u}^{\mathrm{in}}_{0}{}^{-}-{v}^{\mathrm{in}}_{-1}{}^{-}), 
\label{WHeqK}
{{\mathpzc{N}}}({{z}})=\frac{(1+{{z}}){\ellhx}({{z}})^{-1}+1}{3(1-\tfrac{1}{4}{\upomega}^2)}.
\end{eqn}
Using the definition of $\elltg$ in \eqref{WHeqBls4c}, notice that ${{\KK}_{{}}}=({z}\elltg^{-1}+\elltg)/({z}\elltg^{-1}-\elltg)$ which resembles the kernel in \eqref{WHeqBls1}, whereas ${{\KK}}_{{}}$ \eqref{WHeqK} is already analogous to \eqref{WHeqBls0}.

\section{{{WH}} kernel as $2\times2$ matrix: two reducible cases}
\label{matrixprobs1}
In this section, I discuss those scattering problems that involve certain $2\times2$ matrix {{WH}} kernels but which are effectively scalar as they can be reduced to scalar {{WH}} kernels by exploiting a particular kind of disguised symmetry.

First such example is that of scattering due to a semi-infinite crack in triangular lattice \cite{Bls4}.
\begin{figure}[h!]
\centering
\includegraphics[width=\textwidth]{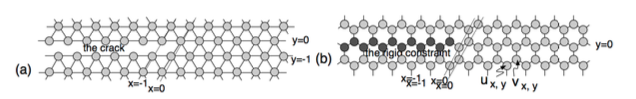}
\caption{Triangular lattice with a semi-infinite row prescribed with discrete Neumann (crack) condition \cite{Bls4,Bls6}, (b) Honeycomb 
with a semi-infinite row having discrete Dirichlet (rigid constraint) \cite{Bls5,Bls6} condition.}
\label{tghxdiffracKC}\end{figure}
It is easily noticable from Fig. \ref{tghxdiffracKC}(a) that there exists no physical symmetry in lattice structure by virtue of the presence of broken slant bonds.
Taking Fourier transform \eqref{unpm} and assuming that ${u}_{\dyp}^{\mathrm{F}}({z})\to 0$ when ${\dyp}\to \pm\infty$, the solution of discrete Helmholtz equation \eqref{discHelmtriang} for scattered field $\su$ is expressed as \eqref{disctriang10} in upper half lattice, while in lower half it is given by (with ${\dyp}>0$)
${u}_{-{\dyp}-1}^{\mathrm{F}}({z})=u_{-1}^{\mathrm{F}}({z})\elltg^{{\dyp}}({z}){z}^{-{\dyp}},$
where the complex function $\elltg$ is given by \eqref{disctrianglambda}.
Throughtout the paper, $H$ is the Heaviside function for integers with $H(x)=0$ for $x<0$ and $H(x)=1$ for $x\ge0$. For the lattice row corresponding to ${\dyp}=0$ and ${\dyp}=-1$, respectively,
in view of structure shown in Fig. \ref{tghxdiffracKC}(a), the discrete Helmholtz equation 
\eqref{discHelmtriang} for the scattered waves is
\begin{eqn}
{u}_{\dxp+1, 0}+{u}_{\dxp-1, 0}+{u}_{\dxp, 1}+{u}_{\dxp-1, 1}+(\tfrac{3}{2}{\upomega}^2-4){u}_{\dxp, 0}-({u}_{\dxp, 0}-{u}_{\dxp, -1})H(\dxp)\\-({u}_{\dxp, 0}-{u}_{\dxp+1, -1})H(\dxp)+({u}^{\mathrm{in}}_{\dxp, 0}-{u}^{\mathrm{in}}_{\dxp, -1})H(-\dxp-1)+({u}^{\mathrm{in}}_{\dxp, 0}-{u}^{\mathrm{in}}_{\dxp+1, -1})H(-\dxp-1)=0,
\end{eqn}
\begin{eqn}
{u}_{\dxp+1, -1}+{u}_{\dxp-1, -1}+{u}_{\dxp+1, -2}+{u}_{\dxp, -2}+(\tfrac{3}{2}{\upomega}^2-4){u}_{\dxp, -1}-({u}_{\dxp, -1}-{u}_{\dxp-1, 0})H(\dxp-1)\\-({u}_{\dxp, -1}-{u}_{\dxp, 0})H(\dxp)+({u}^{\mathrm{in}}_{\dxp, -1}-{u}^{\mathrm{in}}_{\dxp-1, 0})H(-\dxp)+({u}^{\mathrm{in}}_{\dxp, -1}-{u}^{\mathrm{in}}_{\dxp, 0})H(-\dxp-1)=0.
\end{eqn}
Taking the Fourier transform \eqref{unpm} of the above two equations (auxiliary details
 in supplementary 1a), a vector form of {{WH}} equations results which can be written as
\begin{eqn}
\boldsymbol{f}^+({z})+\mathbf{K}({z})\boldsymbol{f}^-({z})=\boldsymbol{c}({z}), {z}\in{\mathcal{A}},
\label{vectorWHeq}
\end{eqn}
\begin{eqn}
\text{where }
\boldsymbol{f}^\pm&=\begin{bmatrix}
u_{0}^{\pm}\\
u_{-1}^{\pm}
\end{bmatrix},
\mathbf{K}^{-1}({z})={N({z})}^{-1}\begin{bmatrix}
N({z})+2&-1-{z}\\
-1-{z}^{-1}&N({z})+2
\end{bmatrix},\\
N({z})&=4-{z}-{z}^{-1}-(1+{z}^{-1})\elltg({z})-\tfrac{3}{2}{\upomega}^2,\\
\boldsymbol{c}({{z}})&=(\mathbf{I}-\mathbf{K}({{z}}))\begin{bmatrix}
{u}^{\mathrm{in}}_{0}{}^{-}({{z}})\\
{u}^{\mathrm{in}}_{-1}{}^{-}({{z}})
\end{bmatrix}+{N({z})}^{-1}{\mathbf{K}({{z}})}
\begin{bmatrix}
-{z} u^{\mathrm{in}}_{0,-1}-{z} u_{0,-1}\\
{u}^{\mathrm{in}}_{0, -1}+{u}_{0,-1}
\end{bmatrix}.
\label{WHeqBls4k}
\end{eqn}
Equivalent to above expression of $N$, it can be seen that $N({z})=(1+{z})\elltg^{-1}({z})-2$ from the equation \eqref{disctrianglambda} for $\elltg$.
Note that $\mathbf{K}$ 
has the structure of Daniele--Khrapkov form \cite{daniele1984solution,Khrapkov2,SpeckPros}
\begin{eqn}
\mathbf{K}({z})=(a_1^2({z})-{z} a_2^2({z}))^{-1}(a_1({z})\mathbf{I}_{2\times 2}+a_2({z})\mathbf{R}({z})), \quad\mathbf{R}({z})=\begin{bmatrix}
0&{z}\\
1&0
\end{bmatrix}, 
\label{Danielefrom}
\end{eqn}
with
$a_1({z})=1+{2}{N({z})}^{-1}, 
a_2({z})={(1+{z}^{-1})}{N({z})}^{-1}.$
Note $\det\mathbf{K}({z})=(a_1^2({z})-{z} a_2^2({z}))^{-1}.$

On the same lines as above case of crack in triangular lattice, with a schematic depiction in Fig. \ref{tghxdiffracKC}(b), I have analyzed the problem of scattering due to a semi-infinite rigid constraint in zigzag honeycomb lattice.
I emphasize that in this case as well there exists no physical symmetry of the physical lattice structure \cite{Bls5} due to the presence of rigid constraint of {\em zigzag} nature \cite{Nakada1996_1,Acik2011}. An equivalent matrix formulation of the problem from \cite{Bls5} is presented in the following.
Notice that there also exists a non-trivial electronic theory behind the proposed difference equations and Dirichlet boundary condition employed \cite{Bls9hx,Bls5c_tube,Bls5ek_tube,Bls5c_tube_media}.
The general solution of \eqref{dHelmholtzhx} (similar to \eqref{dischxsol}) for the scattered wave field $u, v$ in the upper and lower half lattice, respectively,
is 
\begin{eqn}
{v}_{{\dyp}}^{\mathrm{F}}({z})&=v_0^{\mathrm{F}}({z})\ellhx^{{\dyp}}({z}),
{u}_{{\dyp}}^{\mathrm{F}}({z})=v_0^{\mathrm{F}}({z})\frac{1+{z}^{-1}+\ellhx^{-1}}{3(1-\tfrac{1}{4}{\upomega}^2)}\ellhx^{{\dyp}}({z}), {\dyp}>0,\\
{u}_{-{\dyp}}^{\mathrm{F}}({z})&=u_0^{\mathrm{F}}({z})\ellhx^{{\dyp}}({z}){z}^{-{\dyp}},
{v}_{-{\dyp}}^{\mathrm{F}}({z})=u_0^{\mathrm{F}}({z})\frac{1+{z}+{z}\ellhx^{-1}}{3(1-\tfrac{1}{4}{\upomega}^2)}\ellhx^{{\dyp}}({z}){z}^{-{\dyp}}, {\dyp}>0,
\label{dischxsollower}
\end{eqn}
The discrete Helmholtz equation for ${\dyp}=0$ remains same as \eqref{dHelmholtzhx} for the scattered wave field when ${\dxp}\ge0$, however, for $\dxp<0$, the equation needs to be replaced by ${u}^{{t}}_{\dxp, 0}={u}_{\dxp, 0}+{u}^{\mathrm{in}}_{\dxp, 0}=0$, ${v}^{{t}}_{\dxp, 0}={v}_{\dxp, 0}+{v}^{\mathrm{in}}_{\dxp, 0}=0$.
Taking the Fourier transform \eqref{unpm} of these equations 
for $u$ and $v$ at ${\dyp}=0$,
using \eqref{dischxsollower}, it is found that \eqref{vectorWHeq} holds with (details relegated to supplementary 1b)
\begin{eqn}
\boldsymbol{f}^\pm&=\begin{bmatrix}
u_{1}^{\pm}\\
v_{-1}^{\pm}
\end{bmatrix},
\mathbf{K}^{-1}=\mathbf{I}+\mathbf{A}^{-1}, \quad\mathbf{A}({{z}})={{\mathpzc{M}}({z})}^{-1}\begin{bmatrix}-{\mathbb{C}}w&1+{z} \\1+{z}^{-1} &-{\mathbb{C}}w\end{bmatrix},\\
{\mathpzc{M}}({z})&=((1+{z}^{-1})\ellhx({z})+1)/{\mathbb{C}}w, \quad {\mathbb{C}}w=3(1-\tfrac{1}{4}{\upomega}^2),\\
\boldsymbol{c}({{z}})&=(\mathbf{I}-\mathbf{K}({{z}}))\begin{bmatrix}
{u}^{\mathrm{in}}_{1}{}^{-}({{z}})+{z}{u}^{\mathrm{in}}_{0, 0}+{z}{u}_{0, 0}\\
{v}^{\mathrm{in}}_{-1}{}^{-}({{z}})-{v}^{\mathrm{in}}_{-1, 0}-{v}_{-1, 0}
\end{bmatrix}.
\label{WHeqBls5c}
\end{eqn}
The matrix {{WH}} kernel $\mathbf{K}$ can be written in the form \eqref{Danielefrom} 
with
\begin{eqn}
a_1({z})=1-\frac{{\mathbb{C}}w {\mathpzc{M}}({z})}{{\mathbb{C}}w^2-(1+{z})(1+{z}^{-1})}, \quad
a_2({z})=\frac{(1+{z}^{-1}){\mathpzc{M}}({z})}{{\mathbb{C}}w^2-(1+{z})(1+{z}^{-1})}.
\label{DanielefromBls5c}
\end{eqn}

The $2\times2$ matrix {{WH}} kernel $\mathbf{K}$ in \eqref{Danielefrom} can be factorized using an available method \cite{daniele1984solution,Khrapkov2}.
However, the structure of triangular lattice with a semi-infinite crack as well as honeycomb lattice with a semi-infinite zigzag rigid constraint possesses a {\em special} symmetry, which allows a conversion of vector {{WH}} equation \eqref{vectorWHeq} into a scalar {{WH}} equation in both cases. The details of the scalar formulation using alternative choice of triangular lattice coordinates appear in \cite{Bls4} and \cite{Bls5}.
In the context of above expressions \eqref{WHeqBls4k} and \eqref{WHeqBls5c} of the present paper, an equivalent way to state the reduction is that a mapping ${z}\to{z}^2$ allows to capture the special symmetry; as noted in \cite{Bls4} and \cite{Bls5} this construction produces a double degeneracy relative to the fundamental unit cell (Brillouin zone) \cite{Brillouin} as the mentioned map doubly wraps the unit circle. 

\section{More matrix {{WH}} problems on unit circle}
\label{matrixprobs3}
\subsection{}
\begin{figure}[h!]
\includegraphics[width=\textwidth]{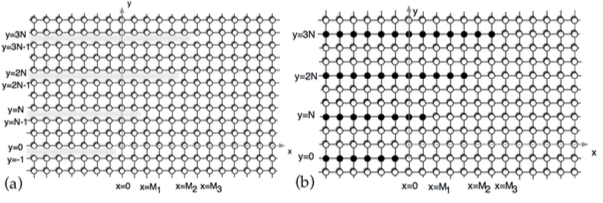}
\caption{Square lattice with $\nu$ number of parallel semi-infinite rows with (a) discrete Neumann (crack) \cite{Bls0,Bls2} and (b) discrete Dirichlet (rigid constraint) condition \cite{Bls1,Bls3}. }
\label{scatteringNcrackslit_sq}\end{figure}
In the continuum formulation of scattering, certain scatterers have been intriguing for some time, for instance, a system of
parallel {S}ommerfeld half-planes \cite{MeisterRottbrand,Meistersys1}.
As a discrete equivalent of such example, associated with a hierarchy of {{WH}} problems, 
consider the scattering due to a finite number of cracks and rigid constraints in square lattice. 
Suppose that the total $\nu$ number of crack tips, as well as $\nu$ rigid constraint tips, end at $\dxp={M}_0,$ $\dxp={M}_1, \dotsc, \dxp={M}_{\nu-1},$ while the $\dyp$ coordinate of the corresponding rows for the former are $\dyp=0+j{N}, -1+j{N},$ and $\dyp=0+j{N}$ for the lattice where $j\in{\mathbb{Z}}_{0}^{{\nu}-1}.$ The notation ${\mathbb{Z}}_{a}^{b}$ stands for the set $\{a, a+1, \dotsc, b\}$.

Using the geometric structure illustrated in Fig. \ref{scatteringNcrackslit_sq}(a) (in the schematic, $\nu=4, {{N}}=3$, ${{M}}_0=0, {{M}}_1=2, {{M}}_2=5, {{M}}_3=7$), for the problem of $\nu$ number of parallel cracks,
the equation of motion at $\dyp=j{{N}}, -1+j{{N}}$ (for $j\in{\mathbb{Z}}_{0}^{{\nu}-1}$) leads to, respectively, for $\dxp\in{\mathbb{Z}},$
${\su}^{{t}}_{\dxp+1, j{{N}}}+{\su}^{{t}}_{\dxp-1, j{{N}}}+{\su}^{{t}}_{\dxp, 1+j{{N}}}
+({\upomega}^2-3){\su}^{{t}}_{\dxp, j{{N}}}+({\su}^{{t}}_{\dxp, -1+j{{N}}}-{\su}^{{t}}_{\dxp, j{{N}}})H(\dxp-{{M}}_{j})=0,$
and
${\su}^{{t}}_{\dxp+1, -1+j{{N}}}+{\su}^{{t}}_{\dxp-1, -1+j{{N}}}+{\su}^{{t}}_{\dxp, -2+j{{N}}}
+({\upomega}^2-3){\su}^{{t}}_{\dxp, -1+j{{N}}}+({\su}^{{t}}_{\dxp, j{{N}}}-{\su}^{{t}}_{\dxp, -1+j{{N}}})H(\dxp-{{M}}_{j})=0.$
Assuming that ${\su}_{\dyp}^{{\mathrm{F}}}\to 0$ when $\dyp\to \pm\infty$, the Fourier transform of the scattered wave field (as complex function, analytic on an annulus 
${{\mathcal{A}}}$)
is found to be
\begin{eqn}
{\su}_{\dyp}^{{\mathrm{F}}}&=\su^{{\mathrm{F}}}_{(\nu-1) {{N}}}{{\lambda}}^{\dyp-{(\nu-1) {{N}}}}~({\dyp} \ge {(\nu-1) {{N}}}),\quad {\su}_{\dyp}^{{\mathrm{F}}}=\su^{{\mathrm{F}}}_{-1}{{\lambda}}^{-({\dyp} +1)}~({\dyp} <0),\\
{\su}_{\dyp+(j-1){{N}}}^{{\mathrm{F}}}
&=f_{\dyp}{\su}_{(j-1){{N}}}^{\mathrm{F}}+f_{{{N}}-1-\dyp}{\su}_{{{{N}}}-1+(j-1){{N}}}^{\mathrm{F}}, f_{\dyp}=\frac{{{\lambda}}^{-2{{{N}}}+2}{{\lambda}}^{\dyp}-{{\lambda}}^{-\dyp}}{{{\lambda}}^{-2{{{N}}}+2}-1}, {\dyp}\in{\mathbb{Z}}_{0}^{{{N}}-1}, j\in{\mathbb{Z}}_{1}^{{\nu}-1}.
\label{finitearraycracksbulk} 
\end{eqn}
Suppose that $\su_{\alpha+j{{N}}}^{\mathrm{F}}({{z}})={z}^{-{{M}}_j}\su_{\alpha+j{{N}}}^{+}({{z}})+{z}^{-{{M}}_j}\su_{\alpha+j{{N}}}^{-}({{z}}),$ with $\su_{\alpha+j{{N}}}^{\pm}({{z}})=\sum\nolimits_{m\in{\mathbb{Z}}^\pm}\su_{m+{{M}}_j,\alpha+j{{N}}}{z}^{-m}$ for $\alpha=0, -1$.
Taking the Fourier transform \eqref{unpm} along the lattice rows $\dyp=j{{N}}, -1+j{{N}}$, 
with
\beqans
\sv_{j{{N}}}^{\mathrm{F}}({{z}})&=&{z}^{-{{M}}_j}\sv_{j{{N}}}^{+}({{z}})+{z}^{-{{M}}_j}\sv_{j{{N}}}^{-}({{z}})=\su_{j{{N}}}^{\mathrm{F}}({{z}})-\su_{-1+j{{N}}}^{\mathrm{F}}({{z}}), \label{FTsemi2}\\
\sv_{j{{N}}}^{\pm}({{z}})&=&\sum\nolimits_{m\in{\mathbb{Z}}^\pm}(\su_{m+{{M}}_j,j{{N}}}-\su_{m+{{M}}_j,-1+j{{N}}}){z}^{-m},\label{FTsemi3}
\eeqans{FTsemi}
it is found that 
\beqans
\hspace{-.1in}(1-{\lambda}^{-1})\su_{(\nu-1){{N}}}^{\mathrm{F}}-(1-f_{-1}){\su}_{-1+(\nu-1){{N}}}^{\mathrm{F}}+f_{{{N}}}{\su}_{(\nu-2){{N}}}^{\mathrm{F}}-2{z}^{-{{M}}_{\nu-1}}(\sv_{(\nu-1){{N}}}^{+}-\sv^{{\mathrm{in}}{-}}_{(\nu-1){{N}}})
=0,\label{finitearraycracks1}\\
(1-{\lambda}^{-1})\su_{(\nu-1){{N}}}^{\mathrm{F}}+(1-f_{-1}){\su}_{-1+(\nu-1){{N}}}^{\mathrm{F}}-f_{{{N}}}{\su}_{(\nu-2){{N}}}^{\mathrm{F}}=0,\label{finitearraycracks2}\\
(1-f_{-1})({\su}_{j{{N}}}^{\mathrm{F}}-\su_{-1+j{{N}}}^{\mathrm{F}})-f_{{{N}}}({\su}_{-1+(j+1){{N}}}^{\mathrm{F}}-{\su}_{(j-1){{N}}}^{\mathrm{F}})-2{z}^{-{{M}}_j}\sv_{j{{N}}}^{+}+2{z}^{-{{M}}_j}\sv^{\mathrm{in}}_{j{{N}}}{}^{-}=0, 
\label{finitearraycracks3}\\
(1-f_{-1})({\su}_{j{{N}}}^{\mathrm{F}}+\su_{-1+j{{N}}}^{\mathrm{F}})-f_{{{N}}}({\su}_{-1+(j+1){{N}}}^{\mathrm{F}}+{\su}_{(j-1){{N}}}^{\mathrm{F}})=0, 
\label{finitearraycracks4}\\
(1-f_{-1})\su_{0}^{\mathrm{F}}-(1-{\lambda}^{-1})\su_{-1}^{\mathrm{F}}-f_{{{N}}}{\su}_{{{{N}}}-1}^{\mathrm{F}}-2{z}^{-{{M}}_0}\sv_{0}^{+}+2{z}^{-{{M}}_0}\sv^{\mathrm{in}}_{0}{}^{-}=0,\label{finitearraycracks5}\\
(1-f_{-1})\su_{0}^{\mathrm{F}}+(1-{\lambda}^{-1})\su_{-1}^{\mathrm{F}}-f_{{{N}}}{\su}_{{{{N}}}-1}^{\mathrm{F}}=0,\label{finitearraycracks6}
\eeqans{finitearraycracks}
(with $j\in{\mathbb{Z}}_{1}^{{\nu}-2}$ for the third and fourth equation, some details relegated to supplementary 2a).
The total $2\nu$ number of equations, \eqref{FTsemi2}${}_2$, \eqref{finitearraycracks2}, \eqref{finitearraycracks4} and \eqref{finitearraycracks6}, can be solved for $\su_{j{{N}}}^{\mathrm{F}}, \su_{-1+j{{N}}}^{\mathrm{F}}$ in terms of $\nu$ entities $\sv_{j{{N}}}^{\mathrm{F}}$, so that the terms containing $\su_{j{{N}}}^{\mathrm{F}}, \su_{-1+j{{N}}}^{\mathrm{F}}$ in \eqref{finitearraycracks1}, \eqref{finitearraycracks3}, \eqref{finitearraycracks5} can be written solely in terms of $\sv_{j{{N}}}^{\mathrm{F}}$ using a matrix, say, denoted by $\mathbf{A}$.
Thus, the coupled {{WH}} equations \eqref{finitearraycracks1}, \eqref{finitearraycracks3}, \eqref{finitearraycracks5} can be written 
as
$(\mathbf{Z}\mathbf{A}\mathbf{Z}^{-1}-2\mathbf{I})\boldsymbol{f}^++\mathbf{Z}\mathbf{A}\mathbf{Z}^{-1}\boldsymbol{f}^-=-2\boldsymbol{f}^{\mathrm{in}}{}^-,$
where
\begin{eqn}
\boldsymbol{f}^\pm=[\sv_{(\nu-1){{N}}}^{\pm}\dotsc\sv_{0{{N}}}^{\pm}]^T,
\boldsymbol{f}^{\mathrm{in}}{}^-=[\sv^{\mathrm{in}}_{(\nu-1){{N}}}{}^{-}\dotsc\sv^{\mathrm{in}}_{0{{N}}}{}^{-}]^T,\quad
\mathbf{Z}=\text{Diag}({z}^{{{M}}_{\nu-1}},\dotsc, {z}^{{{M}}_0}).
\label{arraycracksKfc1}
\end{eqn}
Finally, the {{WH}} can be written as \eqref{vectorWHeq}
with \eqref{arraycracksKfc1} and
\begin{eqn}
\mathbf{K}=\tfrac{1}{2}(\tfrac{1}{2}\mathbf{Z}\mathbf{A}\mathbf{Z}^{-1}-\mathbf{I})^{-1}\mathbf{Z}\mathbf{A}\mathbf{Z}^{-1},\quad
\boldsymbol{c}=(\mathbf{I}-\mathbf{K})\boldsymbol{f}^{\mathrm{in}}{}^-.
\label{arraycracksKfc}
\end{eqn}
In fact, in the special case $\nu=2$
and
$\nu=3,$ respectively, the kernel $\mathbf{K}$ can be simplified to 
\begin{eqn}
\frac{\mathpzc{h}}{\mathpzc{r}}\begin{bmatrix}1&{\lambda}^{{N}} {z}^{{{M}}_1-{{M}}_0} \\
{\lambda}^{{N}} {z}^{{{M}}_0-{{M}}_1}&1\end{bmatrix},\quad
\frac{\mathpzc{h}}{\mathpzc{r}}\begin{bmatrix}1&{\lambda}^{{N}} {z}^{{{M}}_2-{{M}}_1} &{\lambda}^{2{{N}}} {z}^{{{M}}_2-{{M}}_0} \\
{\lambda}^{{N}} {z}^{{{M}}_1-{{M}}_2}&1&{\lambda}^{{N}} {z}^{{{M}}_1-{{M}}_0}\\
{\lambda}^{2{{N}}} {z}^{{{M}}_0-{{M}}_2}&{\lambda}^{{N}} {z}^{{{M}}_0-{{M}}_1}&1
\end{bmatrix},
\label{splkernelK}
\end{eqn}
while 
in general, the components of $\mathbf{K}$ for arbitrary $\nu$ are given by $K_{ij}=\frac{\mathpzc{h}}{\mathpzc{r}}{\lambda}^{|i-j|}{z}^{{{M}}_{\nu-i}-{{M}}_{\nu-j}}$.
Note $\det\mathbf{K}=({\mathpzc{h}}/{\mathpzc{r}})^{\nu}(1-{\lambda}^{2{N}})^{\nu-1}.$

Using the geometric structure illustrated in Fig. \ref{scatteringNcrackslit_sq}(b) (in the schematic, $\nu=4, {{N}}=3$, ${{M}}_0=0, {{M}}_1=2, {{M}}_2=5, {{M}}_3=7$), for the problem of discrete scattering associated with a finite array of semi-infinite rigid constraints,
the equation of motion at $\dyp=j{{N}}$ (for $j\in{\mathbb{Z}}_0^{\nu-1}$) remains \eqref{dHelmholtz} when $\dxp\ge{{M}}_{j}$, whereas
the constraint leads to ${\su}^{{t}}_{\dxp, j{{N}}}=0$ for $\dxp<{{M}}_{j}.$
For this problem of $\nu$ number of parallel rigid constraints, assuming that ${\su}_{\dyp}^{{\mathrm{F}}}\to 0$ when $\dyp\to \pm\infty$, the Fourier transform of the scattered wave field 
is found to be similar to \eqref{finitearraycracksbulk}, i.e.,
\begin{eqn}
{\su}_{\dyp}^{{\mathrm{F}}}&=\su^{{\mathrm{F}}}_{(\nu-1) {{N}}}{{\lambda}}^{\dyp-{(\nu-1) {{N}}}}~({\dyp} \ge {(\nu-1) {{N}}}), \quad{\su}_{\dyp}^{{\mathrm{F}}}=\su^{{\mathrm{F}}}_{0}{{\lambda}}^{-{\dyp}}~({\dyp} \le0),\\
{\su}_{\dyp+(j-1){{N}}}^{{\mathrm{F}}}
&=f_{\dyp}{\su}_{(j-1){{N}}}^{\mathrm{F}}+f_{{{N}}-\dyp}{\su}_{j{{N}}}^{\mathrm{F}}, f_{\dyp}=\frac{{{\lambda}}^{-2{{{N}}}}{{\lambda}}^{\dyp}-{{\lambda}}^{-\dyp}}{{{\lambda}}^{-2{{{N}}}}-1}, {\dyp}\in{\mathbb{Z}}_{0}^{{{N}}}, j\in{\mathbb{Z}}_1^{\nu-1}.
\label{finitearrayrigidbulk}
\end{eqn}
Suppose that
$\su_{j{{N}}}^{\mathrm{F}}={z}^{-{{M}}_j}\su_{j{{N}}}^{+}+{z}^{-{{M}}_j}\su_{j{{N}}}^{-},
\su_{j{{N}}}^{\pm}=\sum\nolimits_{m\in{\mathbb{Z}}^\pm}\su_{m+{{M}}_j,j{{N}}}{z}^{-m}.$
Taking the Fourier transform \eqref{unpm} along the lattice rows $\dyp=j{{N}}$, 
with 
\begin{eqn}
{\mathtt{w}}_{j{{N}}}^{\mathrm{F}}&={z}^{-{{M}}_j}{\mathtt{w}}_{j{{N}}}^{+}+{z}^{-{{M}}_j}{\mathtt{w}}_{j{{N}}}^{-}=\su_{1+j{{N}}}^{\mathrm{F}}+\su_{-1+j{{N}}}^{\mathrm{F}}, \\
{\mathtt{w}}_{j{{N}}}^{\pm}&=\sum\nolimits_{m\in{\mathbb{Z}}^\pm}(\su_{m+{{M}}_j,1+j{{N}}}+\su_{m+{{M}}_j,-1+j{{N}}}){z}^{-m},
\end{eqn}
it is found that (with 
a detailed derivation placed in supplementary 2b)
\beqans
\hspace{-.1in}\hspace{-.2in}(\mathpzc{h}^2+2)( {\mathtt{w}}_{(\nu-1){{N}}}^{+}+{\mathtt{w}}_{(\nu-1){{N}}}^{-})\hspace{-.1in}&=&\hspace{-.1in}({\lambda}+f_1)({\mathtt{w}}_{(\nu-1){{N}}}^{+}-\chinew_{\nu-1})
+\frac{{z}^{{{M}}_{\nu-1}}}{{z}^{{{M}}_{\nu-2}}}f_{{{N}}-1}({\mathtt{w}}_{(\nu-2){{N}}}^{+}-\chinew_{\nu-2}),\label{finitearrayrigid1}\\
(\mathpzc{h}^2+2)({\mathtt{w}}_{j{{N}}}^{+}+{\mathtt{w}}_{j{{N}}}^{-})&=&2f_{1}({\mathtt{w}}_{j{{N}}}^{+}-\chinew_j)+f_{{{N}}-1}{{z}^{{{M}}_j-{{M}}_{j+1}}}({\mathtt{w}}_{(j+1){{N}}}^{+}-\chinew_{j+1})\notag\\
&&+f_{{{N}}-1}{{z}^{{{M}}_j-{{M}}_{j-1}}}({\mathtt{w}}_{(j-1){{N}}}^{+}-\chinew_{j-1}), j\in{\mathbb{Z}}_{1}^{{\nu}-2}\label{finitearrayrigid2}\\
(\mathpzc{h}^2+2)({\mathtt{w}}_{0}^{+}+{\mathtt{w}}_{0}^{-})&=&f_{{{N}}-1}{{z}^{{{M}}_0-{{M}}_{1}}}({\mathtt{w}}_{{{N}}}^{+}-\chinew_{1})+({\lambda}+f_1)({\mathtt{w}}_{0}^{+}-\chinew_0), \label{finitearrayrigid3}\\
\text{with }
\chinew_j&=&(\mathpzc{h}^2+2)\su^{\mathrm{in}}_{j{{N}}}{}^{-}-\su_{{{M}}_j-1,j{{N}}}+{z}\su_{{{M}}_j,j{{N}}}.
\label{defchij}
\eeqans{finitearrayrigid}
Above set of coupled {{WH}} equations \eqref{finitearrayrigid1}, \eqref{finitearrayrigid2}, \eqref{finitearrayrigid3} can be written as a system of equations in the form \eqref{vectorWHeq} with (using \eqref{defchij})
\begin{eqn}
\boldsymbol{f}^\pm=[{\mathtt{w}}_{(\nu-1){{N}}}^{\pm}\dotsc{\mathtt{w}}_{0{{N}}}^{\pm}]^T,\quad
\boldsymbol{c}=(\mathbf{I}-\mathbf{K}){\boldsymbol{\chinew}}, \quad{\boldsymbol{\chinew}}=[\chinew_{(\nu-1)}\dotsc\chinew_{0}]^T.
\end{eqn}
In the special case $\nu=2$ and $\nu=3,$ respectively, the kernel $\mathbf{K}$ can be reduced to the form
\begin{eqn}
\frac{\mathpzc{h}^2+2}{\mathpzc{r}\mathpzc{h}}\begin{bmatrix}1&{\lambda}^{{N}} {z}^{{{M}}_1-{{M}}_0}\\
{\lambda}^{{N}} {z}^{{{M}}_0-{{M}}_1}&1
\end{bmatrix},\quad
\frac{\mathpzc{h}^2+2}{\mathpzc{r}\mathpzc{h}}\begin{bmatrix}1&{\lambda}^{{N}} {z}^{{{M}}_2-{{M}}_1} &{\lambda}^{2{{N}}} {z}^{{{M}}_2-{{M}}_0} \\
{\lambda}^{{N}} {z}^{{{M}}_1-{{M}}_2}&1&{\lambda}^{{N}} {z}^{{{M}}_1-{{M}}_0}\\
{\lambda}^{2{{N}}} {z}^{{{M}}_0-{{M}}_2}&{\lambda}^{{N}} {z}^{{{M}}_0-{{M}}_1}&1
\end{bmatrix}
\label{splkernelC}
\end{eqn}
while,
in general, the components of $\mathbf{K}$ for arbitrary $\nu$ are given by $K_{ij}=\frac{\mathpzc{h}^2+2}{\mathpzc{r}\mathpzc{h}}{\lambda}^{|i-j|}{z}^{{{M}}_{\nu-i}-{{M}}_{\nu-j}}$.
Note $\det\mathbf{K}=(\mathpzc{h}^2+2)^{\nu}/({\mathpzc{r}\mathpzc{h}})^{\nu}(1-{\lambda}^{2{N}})^{\nu-1}.$

The special case of $\nu=2$ stated in \eqref{splkernelK} and \eqref{splkernelC} for a pair of cracks and rigid constraints, respectively, has been studied recently \cite{Bls8pair1,Bls8staggerpair_exact};
the equation for a pair of staggered cracks and rigid constraints has also been solved approximately \cite{GMthesis,Bls8staggerpair_asymp} by asymptotic factorization \cite{Mishuris2014}; see well known continuum counterparts in \cite{Abrahams0,Abrahams2,Abrahams4}.

As $\nu\to\infty$, with ${{M}}_j\equiv {{M}}$ for all $j$, i.e., in the presence of an array of semi-infinite cracks (or rigid constraints) on square lattice, the scattering problem also admits an analysis that can be reduced to scalar {{WH}} facotrization. This appears as a separate article \cite{Bls8arrayfinite} in the same thematic series (on {{WH}} problems) while the original continuum models appeared as early as 50s, see references therein.
Also as ${N}\to\infty$, the kernel $\mathbf{K}$ becomes diagonalized with the entries corresponding to the kernel for individual crack \eqref{WHeqBls0} and rigid constraint \eqref{WHeqBls1}.

\subsection{}
One of the simplest extension of the hierarchy of problems discussed above, and a well known one in continuum formulation \cite{Luneburg1,Rawlins469,HurdLune}, is that of a combination of Neumann and Dirichlet condition on the same plane or planes separated vertically.
As a generalization of a pair of cracks and a pair of rigid constraints, a discrete scattering problem is considered that concerns a crack and a rigid constraint.
Both kinds of problems are described in the following.

\begin{figure}[!ht]
\centering
\includegraphics[width=\textwidth]{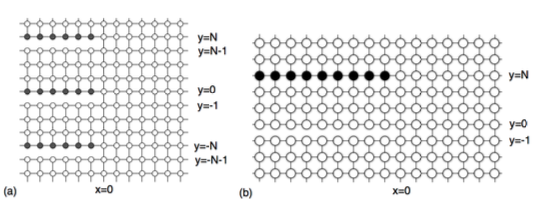}
\caption[ ]{(a) Discrete analogue of scattering due to half-planes with alternate placement of Neumann and Dirichlet condition \cite{Luneburg1}.
(b) Discrete analogue of scattering by hard and soft parallel half-planes \cite{HurdLune}.}
\label{sqdiffracmixed2}
\end{figure}

I consider the array of half-row defects with separation by ${{N}}$ as shown in Fig. \ref{sqdiffracmixed2}(a), i.e.,
suppose that $\su^{{t}}_{\dxp,0+j{N}}=0$ for $\dxp<0$ while the bonds between $\dyp=0+j{N}$ and $\dyp=-1+j{N}$ are also broken. Thus the equation of motion at $\dyp=0+j{{N}}$ and $\dyp=-1+j{{N}}$ remains \eqref{dHelmholtz} for $\dxp\ge0$, whereas
for $\dyp=0+j{{N}}, \dxp<0$ ${\su}_{\dxp, {{N}}}+{\su}^{\mathrm{in}}_{\dxp, {{N}}}=0,$ and for $\dyp=-1+j{{N}}, \dxp<0$,
\begin{eqn}
\su^{{t}}_{\dxp+1,\dyp}+\su^{{t}}_{\dxp-1,\dyp}+\su^{{t}}_{\dxp,\dyp+1}+\su^{{t}}_{\dxp,\dyp-1}H(\dxp)+({\upomega}^2-3)\su^{{t}}_{\dxp,\dyp}-\su^{{t}}_{\dxp,\dyp}H(\dxp)=0.
\label{crackcondn}
\end{eqn}
The Fourier transform \eqref{unpm} of 
the equation at
$\dyp=0+j{{N}}$ and $\dyp=-1+j{{N}}$ leads to, respectively,
\begin{eqn}
-(\mathpzc{h}^2+2)\su_{0+j{N}}^{\mathrm{F}}+\su_{-1,0+j{N}}-{z}\su_{0,0+j{N}}+\su_{1+j{N}}^{+}+\su_{-1+j{N}}^{+}=-(\mathpzc{h}^2+2)\su_{0+j{N}}^{-},\\
-(\mathpzc{h}^2+1)\su_{-1+j{N}}^{\mathrm{F}}+\su_{-2+j{N}}^{\mathrm{F}}+\su_{0+j{N}}^{+}-\su_{-1+j{N}}^{+}=\su^{\mathrm{in}}_{0+j{N}}{}^{-}-\su^{\mathrm{in}}_{-1+j{N}}{}^{-}.
\label{listofeqns}
\end{eqn}
By applying the Floquet--Bloch theorem to the lattice structure interacting with the incident wave \eqref{uinc}, similar to \cite{Bls8arrayfinite}, 
\begin{eqn}
\su_{\dxp, \dyp+j{N}}=\phasefac^j\su_{\dxp, \dyp}, \dyp\in{\mathbb{Z}}_0^{{N}-1},\quad\text{where }\phasefac=e^{-i{\upkappa}_y{N}}.
\end{eqn}
Moreover, (recall \eqref{finitearraycracksbulk}) $\su_{\dyp}^{\mathrm{F}}=f_{\dyp}\su_{0}^{\mathrm{F}}+f_{{N}-1-\dyp}\su_{{N}-1}^{\mathrm{F}},$ $f_{\dyp}=\frac{{{\lambda}}^{-2{{{N}}}+2}{{\lambda}}^{\dyp}-{{\lambda}}^{-\dyp}}{{{\lambda}}^{-2{{{N}}}+2}-1},$ so that finally
\eqref{listofeqns} can be written as \eqref{vectorWHeq} 
where 
\begin{eqn}
\hspace{-.1in}\mathbf{K}
=(P_{{N}}Q_{{N}})^{-1}\begin{bmatrix}
-\frac{\phasefac^{-1}P_{{N}}+{\lambda}^{{N}+2}Q_{{N}}}{1-{\lambda}^2}&\frac{-{\lambda}^{{N}-1}P_{{N}}+{\lambda}^{2}\phasefac Q_{{N}}}{(1+{\lambda})}\\
{\lambda}\frac{-\phasefac^{-1}P_{{N}-1}+{\lambda}^{{N}}Q_{{N}-1}}{(1+{\lambda})}&\frac{(1-{\lambda})}{(1+{\lambda})}(1-{\lambda}^{2{N}})
\end{bmatrix}, P_{{N}}={\lambda}^{{N}}-\phasefac, Q_{{N}}={\lambda}^{{N}}-\phasefac^{-1},\\
\boldsymbol{f}_\pm=\begin{bmatrix}
\su_{1}^{\pm}\\
\sv_{0}^{\pm}
\end{bmatrix},\quad
\boldsymbol{c}({z})=(\mathbf{I}-\mathbf{K}({z}))\begin{bmatrix}
1&1\\
0&1
\end{bmatrix}\begin{bmatrix}
(\mathpzc{h}^2+1)\su^{\mathrm{in}}_{0}{}^{-}({z})-\su_{-1,0}+{z}\su_{0,0}\\
\sv^{\mathrm{in}}_{0}{}^{-}({z})
\end{bmatrix}.
\end{eqn}
As ${N}\to\infty,$ using ${\lambda}^{N}\to0$,
the kernel for a single semi-infinite defect is found (discrete anagloue of classical problem \cite{buyukaksoy1995note}).
Note $\det\mathbf{K}=(1+{\lambda}^3) ({\lambda}^{-{N}}+ {\lambda}^{{N}-1})/((1 + {\lambda})^2({\lambda}^{-{N}} +{\lambda}^{{N}}- (\phasefac + \phasefac^{-1}))).$

Consider a crack and a rigid constraint separated by ${{N}}$ as depicted in Fig. \ref{sqdiffracmixed2}(b). Then for $\dxp<0$, $\su^{{t}}_{\dxp,{N}}=0$ while the bonds between $\dyp=0$ and $\dyp=-1$ are broken. Thus the equation of motion at $\dyp={N}, 0$ and $-1$ remains \eqref{dHelmholtz} for $\dxp\ge0$, whereas
for $\dyp={N}, \dxp<0$ ${\su}_{\dxp, {{N}}}+{\su}^{\mathrm{in}}_{\dxp, {{N}}}=0,$ for $\dyp=0, \dxp<0$, $\su^{{t}}_{\dxp+1,\dyp}+\su^{{t}}_{\dxp-1,\dyp}+\su^{{t}}_{\dxp,\dyp-1}+\su^{{t}}_{\dxp,\dyp+1}H(\dxp)+({\upomega}^2-3)\su^{{t}}_{\dxp,\dyp}-\su^{{t}}_{\dxp,\dyp}H(\dxp)=0,$ and for $\dyp=-1, \dxp<0$, \eqref{crackcondn} holds.
The Fourier transform \eqref{unpm} of the equation of motion at $\dyp={{N}}$ and $\dyp=0, -1$ leads to, respectively,
\begin{eqn}
-(\mathpzc{h}^2+2)\su_{{{N}}}^{\mathrm{F}}+\su_{-1,{{N}}}-{z}\su_{0,{{N}}}+\su_{{{N}}+1}^{+}+\su_{{{N}}-1}^{+}=-{\upomega}^2(\su_{{{N}}}^{-}+\su^{\mathrm{in}}_{{{N}}}{}^{-})-(\mathpzc{h}^2+2)\su_{{{N}}}^{-},\\
-(\mathpzc{h}^2+1)\su_{0}^{\mathrm{F}}+\su_{1}^{\mathrm{F}}-\sv_{0}^{+}=-\sv^{\mathrm{in}}_{0}{}^{-}, \quad -(\mathpzc{h}^2+1)\su_{-1}^{\mathrm{F}}+\su_{-2}^{\mathrm{F}}+\sv_{0}^{+}=\sv^{\mathrm{in}}_{0}{}^{-}.
\label{pairCK}
\end{eqn}
It is easy to see that (recall \eqref{finitearrayrigidbulk})
$\su_{\dyp}=f_{\dyp}\su_{0}^{\mathrm{F}}+f_{{{N}}-\dyp}\su_{{{N}}}^{\mathrm{F}}, f_{\dyp}=(\frac{{{\lambda}}^{-2{{{N}}}}{{\lambda}}^{\dyp}-{{\lambda}}^{-\dyp}}{{{\lambda}}^{-2{{{N}}}}-1}).$
Let
$\sv_{0}^{\mathrm{F}}=\su_{0}^{\mathrm{F}}-\su_{-1}^{\mathrm{F}}$
and
${\mathtt{w}}_{{{N}}}^{\mathrm{F}}=\su_{{{N}}+1}^{F}+\su_{{{N}}-1}^{F}=(f_{{{N}}-1}\su_{0}^{\mathrm{F}}+f_{1}\su_{{{N}}}^{\mathrm{F}})+{\lambda}\su_{{{N}}}^{\mathrm{F}}$; the latter gives $\su_{{{N}}}^{\mathrm{F}}=(f_1+{\lambda})^{-1}({\mathtt{w}}_{{{N}}}^{\mathrm{F}}-f_{{{N}}-1}\su_{0}^{\mathrm{F}})$.
Hence, \eqref{pairCK} can be simplified to the {{WH}} equation \eqref{vectorWHeq} with
\begin{eqn}
\mathbf{K}=\begin{bmatrix}
\frac{1+{\lambda}^2}{1-{\lambda}^2}&\frac{-{\lambda}^{{N}}(1+{\lambda}^2)}{1+{\lambda}}\\
\frac{{\lambda}^{{{N}}+1}}{1+{\lambda}}&\frac{1-{\lambda}}{1+{\lambda}}
\end{bmatrix},
\boldsymbol{f}^\pm=\begin{bmatrix}
{\mathtt{w}}_{{{N}}}^{\pm}\\
\sv{0}^{\pm}
\end{bmatrix},
\boldsymbol{c}=(\mathbf{I}-\mathbf{K})\begin{bmatrix}
(\mathpzc{h}^2+2)\su^{\mathrm{in}}_{{{N}}}{}^{-}-\su_{-1,{{N}}}+{z}\su_{0,{{N}}}\\
\sv^{\mathrm{in}}_{0}{}^{-}
\end{bmatrix}.
\label{pairCKfns}
\end{eqn}
As ${N}\to\infty$, the kernel $\mathbf{K}$ becomes diagonalized with the entries corresponding to the kernel for individual crack \eqref{WHeqBls0} and rigid constraint \eqref{WHeqBls1}.
Note $\det\mathbf{K}=(1+{\lambda}^2) (1+ {\lambda}^{2{N}+1})/(1 + {\lambda})^2.$

\begin{figure}[h!]
\centering
\includegraphics[width=\textwidth]{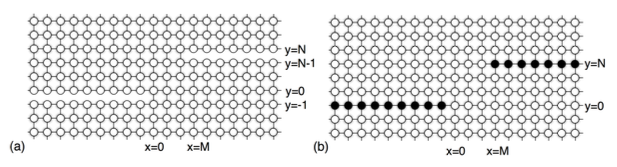}
\caption{Square lattice with a pair of opposite facing but parallel semi-infinite rows with (a) discrete Neumann (crack) \cite{Bls0,Bls2} and (b) discrete Dirichlet (rigid constraint) condition \cite{Bls1,Bls3}.}
\label{scatteringoppositecrackslit_sq}\end{figure}

\subsection{}
Consider the problem of scattering due to opposing tips of cracks or rigid constraints; see Fig. \ref{scatteringoppositecrackslit_sq}. These problems can be considered discrete analogues of continuum problem \cite{Kashyap,Abrahams3,Daniele2018}.
Consider first the case of cracks and then rigid constraints. 
Using the geometric structure illustrated in Fig. \ref{scatteringoppositecrackslit_sq}(a) (in the schematic, ${{N}}=3$ and ${{M}}=3$),
the equation of motion at $\dyp={{N}}, {{N}}-1, 0, -1,$ respectively, implies
\begin{eqn}
{\su}^{{t}}_{\dxp+1, {{N}}}+{\su}^{{t}}_{\dxp-1, {{N}}}+{\su}^{{t}}_{\dxp, {{N}}+1}
+({\upomega}^2-3){\su}^{{t}}_{\dxp, {{N}}}+({\su}^{{t}}_{\dxp, {{N}}-1}-{\su}^{{t}}_{\dxp, {{N}}})H(-\dxp+{{M}}-1)=0,\\
{\su}^{{t}}_{\dxp+1, {{N}}-1}+{\su}^{{t}}_{\dxp-1, {{N}}-1}+{\su}^{{t}}_{\dxp, {{N}}-2}
+({\upomega}^2-3){\su}^{{t}}_{\dxp, {{N}}-1}+({\su}^{{t}}_{\dxp, {{N}}}-{\su}^{{t}}_{\dxp, {{N}}-1})H(-\dxp+{{M}}-1)=0,\\
{\su}^{{t}}_{\dxp+1, 0}+{\su}^{{t}}_{\dxp-1, 0}+{\su}^{{t}}_{\dxp, 1}
+({\upomega}^2-3){\su}^{{t}}_{\dxp, 0}+({\su}^{{t}}_{\dxp, -1}-{\su}^{{t}}_{\dxp, 0})H(\dxp)=0,\\
{\su}^{{t}}_{\dxp+1, -1}+{\su}^{{t}}_{\dxp-1, -1}+{\su}^{{t}}_{\dxp, -2}
+({\upomega}^2-3){\su}^{{t}}_{\dxp, -1}+({\su}^{{t}}_{\dxp, 0}-{\su}^{{t}}_{\dxp, -1})H(\dxp)=0.
\end{eqn}
Taking the Fourier transform \eqref{unpm} along 
$\dyp={{N}}, {{N}}-1, 0, -1,$ it is found that,
respectively,
\begin{eqn}
(1-{\lambda}^{-1})\su_{{{N}}}^{\mathrm{F}}-(1-f_{-1})\su_{{{N}}-1}^{\mathrm{F}}+f_{{{N}}}\su_{0}^{\mathrm{F}}-2{z}^{-{{M}}}\sv_{{{N}}}^{-}&=-2{z}^{-{{M}}}\sv^{\mathrm{in}}_{{{N}}}{}^{+},\\
(1-f_{-1})\su_{0}^{\mathrm{F}}-f_{{{N}}}\su_{{{N}}-1}^{\mathrm{F}}-(1-{\lambda}^{-1})\su_{-1}^{\mathrm{F}}-2\sv_{0}^{+}&=-2\sv^{\mathrm{in}}_{0}{}^{-},\\
(1-{\lambda}^{-1})\su_{{{N}}}^{\mathrm{F}}+(1-f_{-1})\su_{{{N}}-1}^{\mathrm{F}}-f_{{{N}}}\su_{0}^{\mathrm{F}}&=0,\\
(1-f_{-1})\su_{0}^{\mathrm{F}}-f_{{{N}}}\su_{{{N}}-1}^{\mathrm{F}}+(1-{\lambda}^{-1})\su_{-1}^{\mathrm{F}}&=0,
\label{opposingcracks}
\end{eqn}
where (recall \eqref{finitearraycracksbulk}) $\su_{\dyp}^{\mathrm{F}}=f_{\dyp}\su_{0}^{\mathrm{F}}+f_{{N}-1-\dyp}\su_{{N}-1}^{\mathrm{F}},$ $f_{\dyp}=\frac{{{\lambda}}^{-2{{{N}}}+2}{{\lambda}}^{\dyp}-{{\lambda}}^{-\dyp}}{{{\lambda}}^{-2{{{N}}}+2}-1}.$
Hence, \eqref{vectorWHeq} follows with 
\begin{eqn}
\mathbf{K}=\begin{bmatrix}\frac{1+{\lambda}}{1-{\lambda}}&{\lambda}^{{N}} {z}^{{{M}}} \\
-{\lambda}^{{N}} {z}^{-{{M}}}&\frac{1-{\lambda}}{1+{\lambda}}(1-{\lambda}^{2{{N}}})\end{bmatrix},
\boldsymbol{f}^\pm=\begin{bmatrix}\sv_{{{N}}}^{\pm}\\\sv_{0}^{\pm}\end{bmatrix},\quad
\boldsymbol{c}=(\mathbf{I}-\mathbf{K})(\mathbf{I}_2-\mathbf{I}_1)
\begin{bmatrix}
\sv^{\mathrm{in}}_{{{N}}}{}^{+}\\
\sv^{\mathrm{in}}_{0}{}^{-}
\end{bmatrix},
\label{opposingcrackeq}
\end{eqn}
\begin{eqn}
\text{where }
\mathbf{I}_1=\begin{bmatrix}1&0\\0&0\end{bmatrix}, \mathbf{I}_2=\begin{bmatrix}0&0\\0&1\end{bmatrix}.
\label{I1I2}
\end{eqn}

Using the geometric structure illustrated in Fig. \ref{scatteringoppositecrackslit_sq}(b),
the equation of motion at $\dyp={{N}}, \dxp\le{{M}}-1$ and $\dyp=0, \dxp\ge0$ remains \eqref{dHelmholtz}, whereas
for $\dyp={{N}}, \dxp>{{M}}-1$ and $\dyp=0, \dxp<0$, 
${\su}_{\dxp, {{N}}}+{\su}^{\mathrm{in}}_{\dxp, {{N}}}=0.$ 
Now (recall \eqref{finitearrayrigidbulk})
${\su}_{\dyp}^{{\mathrm{F}}}=f_{\dyp}{\su}_{0}^{\mathrm{F}}+f_{{{N}}-\dyp}{\su}_{{{N}}}^{\mathrm{F}}, f_{\dyp}=\frac{{{\lambda}}^{-2{{{N}}}}{{\lambda}}^{\dyp}-{{\lambda}}^{-\dyp}}{{{\lambda}}^{-2{{{N}}}}-1}, {\dyp}\in{\mathbb{Z}}_{0}^{{{N}}}.$
Taking the Fourier transform \eqref{unpm} along the lattice rows $\dyp={{N}}$ and $\dyp=0$, it is found that 
\begin{eqn}
(\mathpzc{h}^2+2)({\mathtt{w}}_{{{N}}}^{+}+{\mathtt{w}}_{{{N}}}^{-})=({\lambda}+f_1)({\mathtt{w}}_{{{N}}}^{-}-\chinew_{{N}})+{z}^{{{M}}}f_{{{N}}-1}({\mathtt{w}}_{0}^{+}-\chinew_{0}),\\
(\mathpzc{h}^2+2)({\mathtt{w}}_{0}^{+}+{\mathtt{w}}_{0}^{-})=f_{{{N}}-1}{{z}^{-{{M}}}}({\mathtt{w}}_{{{N}}}^{-}-\chinew_{{{N}}})+({\lambda}+f_1)({\mathtt{w}}_{{{N}}}^{+}-\chinew_0),\\
\text{where }
\chi_{{{N}}}=(\mathpzc{h}^2+2)\su^{\mathrm{in}}_{{{N}}}{}^{+}+{z}^{{{M}}}(\su_{{{M}}-1,{{N}}}-{z}\su_{{{M}},{{N}}}), \chi_{0}=(\mathpzc{h}^2+2)\su^{\mathrm{in}}_{0}{}^{-}-\su_{-1,0}+{z}\su_{0,0}.
\label{opposingsliteq}
\end{eqn}
After simplification of the coefficient matrices, above can be written as the {{WH}} equation \eqref{vectorWHeq} with (recall \eqref{I1I2})
\begin{eqn}
\mathbf{K}=\begin{bmatrix}\frac{1-{\lambda}^2}{1+{\lambda}^2}&{\lambda}^{{N}} {z}^{{{M}}} \\
-{\lambda}^{{N}} {z}^{-{{M}}}&\frac{1+{\lambda}^2}{1-{\lambda}^2}(1-{\lambda}^{2{{N}}})\end{bmatrix},
\boldsymbol{f}^\pm=\begin{bmatrix}{\mathtt{w}}_{{{N}}}^{\pm}\\{\mathtt{w}}_{0}^{\pm}\end{bmatrix}, \quad
\boldsymbol{c}=(\mathbf{I}-\mathbf{K})(\mathbf{I}_2-\mathbf{I}_1){\boldsymbol{\chinew}},
\label{kerneloppslits}
\end{eqn}
where ${\boldsymbol{\chinew}}=[\chinew_{{{N}}},\chinew_{0}]^T$ (following \eqref{defchij} with ${{M}}_0=0$).

On the lines of another classical problem \cite{buyukaksoy1994plane}, consider the scenario when the two different kinds of defects are placed in the geometry of Fig. \ref{scatteringoppositecrackslit_sq}, say, a crack at $\dxp<0$ between $\dyp=0$ and $\dyp=-1$ while a rigid constraint at $\dxp\ge{M}$ and $\dyp={N}$.
In this case, the {{WH}} equation \eqref{vectorWHeq} is found with ($\chinew_{{{N}}}$ defined by \eqref{opposingsliteq})
\begin{eqn}
\mathbf{K}=\begin{bmatrix}
\frac{1-{{\lambda}}^2}{1 + {{\lambda}}^2}& - {(1-{{\lambda}})}{{\lambda}}^{{{N}}}{z}^{{M}} \\
- \frac{(1-{{\lambda}}){\lambda}}{(1 + {{\lambda}}^2) }{{\lambda}}^{ {{N}}}{z}^{-{{M}}} & \frac{1-{{\lambda}}}{1+{{\lambda}}}\left(1+{{\lambda}}^{2 {{N}}+1}\right)
\end{bmatrix},
\boldsymbol{f}^\pm=\begin{bmatrix}{\mathtt{w}}_{{{N}}}^{\pm}\\\sv_{0}^{\pm}\end{bmatrix}, 
\boldsymbol{c}=(\mathbf{I}-\mathbf{K})(\mathbf{I}_2-\mathbf{I}_1)\begin{bmatrix}
\chinew_{{{N}}}\\
\sv^{\mathrm{in}}_{0}{}^{-}
\end{bmatrix}.
\label{kerneloppKC}
\end{eqn}
Remarkably enough, note that $\det\mathbf{K}=1$ in \eqref{opposingcrackeq} and \eqref{kerneloppslits}, but $\det\mathbf{K}=(1-{\lambda})^2/(1+{\lambda}^2)$ in \eqref{kerneloppKC}.
As ${N}\to\infty,$ the kernels for a single semi-infinite crack and rigid constraint are found.

\section{Concluding remarks}
\label{discussn}
There are many applications of discrete {{WH}} method to scattering of waves by edges in lattice structures. I have solved some of these problems during the recent few years but a lot of interesting problems remain to be tackled. I have included in this paper a set of, so far, unsolved problems in third section. Besides the questions related to the {{WH}} factorization there also exist several other aspects, for instance, the generalizations from semi-infinite defects to finite defects \cite{Bls2,Bls3,Kapanadze2018}, issues of the connection with continuum problems \cite{Bls2,Bls31}, analysis based on Toeplitz operator theory \cite{Silbermann,Bls2,Bls3}
using the lattice Green's function, effect of confinements \cite{Bls9s,Bls12s}, questions of existence and uniqueness of the solution (some relevant results appear in \cite{Bls2,Bls3} as well as \cite{Bls6}) with damping as well as without damping, vector {{WH}} problems scattering involving for in-plane motion.
A set of matrix {{WH}} problems of structure similar to that presented in third section also arise in the problem of 
scattering from a crack tip with a damage zone \cite{Bls32},
where the analysis of \cite{Bls0,Bls2} is furthered
from sharp to cohesive crack tip, but 
at the same time makes an interesting connect with a formidable factorisation problem in matrix {{WH}} kernels. 

\section*{Acknowledgments}
I thank SERB MATRICS grant MTR/2017/000013 for the support. 

\renewcommand*{\bibfont}{\footnotesize}
\printbibliography

\end{document}